\documentclass[11pt]{article}
\usepackage{amsmath,epsf,amssymb,latexsym,enumerate,shadow,array,color,bbold}
\usepackage[nosort]{cite}
\usepackage{slashed}
\usepackage{graphicx,wasysym}

\DeclareFontFamily{U}{rsf}{} \DeclareFontShape{U}{rsf}{m}{n}{
  <5> <6> rsfs5 <7> <8> <9> rsfs7 <10-> rsfs10}{}
\DeclareMathAlphabet\Scr{U}{rsf}{m}{n} \makeatletter
\@addtoreset{equation}{section} \makeatother

\def\be{\begin{equation}}
\def\ee{\end{equation}}
\def\ba{\begin{array}}
\def\ea{\end{array}}

\newcommand{\bea}{\begin{eqnarray}}
\newcommand{\eea}{\end{eqnarray}}
\def\beqa{\begin{eqnarray}}
\def\eeqa{\end{eqnarray}}
\def\NN{{\cal N}}
\def\aD3{{\overline {\rm D3}}}
\def\IZ{{\bf Z}}
\def\IC{{\bf C}}
\def\IS{{\bf S}}
\def\IT{{\bf T}}
\def\tr{{\rm tr \,}}

\def\R{\left(T+T^*\right)}

\newcommand{\drawsquare}[2]{\hbox{%
\rule{#2pt}{#1pt}\hskip-#2pt
\rule{#1pt}{#2pt}\hskip-#1pt
\rule[#1pt]{#1pt}{#2pt}}\rule[#1pt]{#2pt}{#2pt}\hskip-#2pt
\rule{#2pt}{#1pt}}

\newcommand{\fund}{\raisebox{-.5pt}{\drawsquare{6.5}{0.4}}}
\newcommand{\antifund}{\overline{\fund}}

\def\rmi{{\rm i}}
\def\rmd{{\rm d}}

\newsavebox{\uuunit}
\sbox{\uuunit}
    {\setlength{\unitlength}{0.825em}
     \begin{picture}(0.6,0.7)
        \thinlines
        \put(0,0){\line(1,0){0.5}}
        \put(0.15,0){\line(0,1){0.7}}
        \put(0.35,0){\line(0,1){0.8}}
       \multiput(0.3,0.8)(-0.04,-0.02){12}{\rule{0.5pt}{0.5pt}}
     \end {picture}}

\usepackage{ifpdf}
\ifx\pdfoutput\undefined
   \pdffalse
\else
   \pdfoutput=1
   \pdftrue
  \usepackage[pdftex]{hyperref}
\def\u0{{\underline 0}}

\def\url{{\underline {r+\ell}}}

\newcommand{\rf}[1]{(\ref{#1})}

\def\rmi{{\rm i}}
\def\rmd{{\rm d}}

\begin{document}

\begin{titlepage}

\rightline{DAMTP-2015-36}
\rightline{IFT-UAM/CSIC-15-079}

\hskip 0.5cm

\vskip 0.5cm

\begin{center}

{\LARGE{\bf String Theory Realizations of the Nilpotent Goldstino   }}

\

\

 { \bf  {  Renata Kallosh$^1$}, Fernando Quevedo$^{2,3}$  and Angel M. Uranga $^4$},
{\small\sl\noindent

\

$^1$ {\it Department of Physics and SITP, Stanford University, Stanford, CA
94305 USA} \\\smallskip
$^2$ {\it ICTP, Strada Costiera 11, 34151 Trieste,  Italy}\\\smallskip
$^3$ {\it DAMTP, CMS, University of Cambridge, Wilberforce Road, Cambridge, CB3 0WA, UK}\\\smallskip
$^4$ {\it Instituto de F\'{\i}sica Te\'orica UAM-CSIC, c/ Nicol\'as Cabrera 13-15, 28049 Madrid, Spain}\\\smallskip
 }

\end{center}

Email:  \texttt{kallosh@stanford.edu, f.quevedo@damtp.cam.ac.uk, angel.uranga@uam.es}

\vskip 1 cm

\begin{abstract}

We describe in detail how the spectrum of a single anti-D3-brane in four-dimensional orientifolded  IIB string models reproduces precisely  the field content of a  
nilpotent chiral superfield with the only  physical component corresponding to the fermionic goldstino. In particular we explicitly consider  a single anti-D3-brane  on top of an O3-plane in warped throats, induced by $(2,1)$ fluxes. More general systems including several anti-branes and other orientifold planes are also discussed. This provides further evidence to the claim that non-linearly realized supersymmetry due to the presence of 
antibranes in string theory can be described by supersymmetric theories including nilpotent superfields. Implications to the KKLT and related scenarios of de Sitter moduli stabilization,  to cosmology and to the structure of soft SUSY-breaking terms are briefly discussed.

\end{abstract}

\vspace{24pt}
\end{titlepage}

\tableofcontents

\newpage

\section{Introduction}

It is well known that the presence of anti-branes in otherwise supersymmetric string configurations breaks supersymmetry. Describing this effect in a properly defined effective field theory is an interesting challenge. In particular,
the KKLT scenario of de Sitter moduli stabilisation \cite{Kachru:2003aw,Kachru:2003sx}  relies on the presence of at least one anti-D3-brane ($ \aD3 $)  to lift the supersymmetric AdS minimum and allow the possibility of dS string vacua. The uplift is due to the positive energy provided by the tension of the $ \aD3 $ brane located at the tip of a warped throat. 

Even though it is generally agreed that the presence of an antibrane breaks supersymmetry spontaneously, see for example  \cite{Polchinski:1998rr}, 
a manifestly supersymmetric action describing this effect was missing until recently.  The corresponding action of the $ \aD3 $ was presented  recently in \cite{Bergshoeff:2015jxa} starting from a single $\kappa$-symmetric brane in the supersymmetric background with fluxes.  Using the consistent supersymmetric orientifold condition for the fields on the brane one finds that the vectors and scalars are cut off in this procedure. It corresponds to placing the $ \aD3 $ on top of an O3-plane,  and the surviving part of the  brane action coincides with the Volkov-Akulov (VA) action \cite {Volkov:1973ix}. This action  has a  non-linearly realized supersymmetry on a single ${\cal N}=1$  fermionic goldstino which has no bosonic supersymmetric partners.  The Volkov-Akulov goldstino model has also an alternative description via a nilpotent chiral multiplet \cite{rocek,Komargodski:2009rz}. In such a multiplet  the scalar component, sgoldstino, is not a fundamental field but a bilinear combination of the fermions.  The auxiliary field of the nilpotent multiplet is not vanishing, which signifies a spontaneously broken supersymmetry.

The renewed interest to KKLT construction of de Sitter vacua is partly due to improved observational data on dark energy and inflationary cosmology. The update on  dark energy follows from combining Planck data with other astrophysical data, including Type Ia supernovae.  The equation of state of dark energy is now, according to \cite{Ade:2015xua}
\be
w = - 1.006 \pm  0.045 \, .
\ee
This  supports the idea behind the KKLT construction and other constructions such as the large volume scenario (LVS) \cite{lvs}\  that lead to the string landscape scenario, that a cosmological constant with $w=-1$ remains a good fit to data.  In fact it is a much better fit than the one in 2003 when this construction was suggested \footnote{For other approaches towards de Sitter space
in string compactifications see \cite{susyds0,susyds1,susyds2,susyds3,Achucarro:2007qa,Kallosh:2014oja}.}.

Further motivations for  nilpotent superfields come from cosmology. The recent bottom-up approach to cosmology \cite{Ferrara:2014kva, Kallosh:2014via,Carrasco:2015pla} using an effective d=4 ${\cal N}=1$ supergravity has  very nice phenomenological features. Namely, new supergravity models were constructed depending on two chiral superfields \cite{Ferrara:2014kva}, an inflaton superfield  and a nilpotent superfield $X$ satisfying the nilpotency condition $X^2(x, \theta)=0$. 
These models agree nicely with the Planck data \cite{Ade:2015xua},  during inflation the scale of ${\delta \rho\over \rho}$ and the tilt of a power spectrum $n_s$ take their known observational 
values. 
Meanwhile, the level of primordial gravity waves $r$ depends on the curvature of the moduli space and is therefore flexible with regard to future discovery of gravity waves or a new bound on $r$.  At the minimum of inflationary potential in the recent models in \cite{Carrasco:2015pla}  supersymmetry is broken spontaneously in de Sitter vacua and the cosmological constant is given by \footnote{We write the scalar potential and then the cosmological constant in units of $M_{planck}$ so there is an implicit $M_{planck}^2$ factor on  the right hand side of the equation.}
\be
\Lambda= M^2- 3 m_{3/2}^2 \, .
\ee
 Here $M$,  the scale of supersymmetry breaking by goldstino at the minimum,   is the value of the auxiliary field of the nilpotent multiplet. This SUSY breaking scale $M$ in the context of models in \cite{Carrasco:2015pla} will be somehow restricted by the LHC discovery/non-discovery of  supersymmetry, depending on how it appears in the soft-breaking terms.  When the universal goldstino contribution to energy
$M^2$  is slightly larger than the one from gravitino $-3 m^2_{3/2}$, the models describe the dark energy with a cosmological constant $\Lambda$ in the spirit of the string landscape.

A complete self-contained proof of  non-linearly realized supersymmetry for the single $ \aD3 $ on top of the O3 plane is contained in \cite{Bergshoeff:2015jxa}. The result has been derived using a 
powerful and rigorous method of a local fermionic $\kappa$-symmetry where the local supergravity background is represented by a superspace with fermionic coordinates. However, there are certain issues that benefit from a direct string theory analysis. For instance, it is interesting to provide a direct microscopic computation of the worldvolume field content of  $\aD3$-branes on top of orientifold planes, by using standard open string worldsheet techniques. In addition, it is not a priori obvious that it is possible to introduce O3-planes at the bottom of warped throats. Even though this may seem a question of finding suitable supergravity backgrounds, it turns out that this quest is best approached by using holography to construct field theories whose RG flow produces the desired orientifolded warped throat, generalizing the KS construction \cite{Klebanov:2000hb}.
 
We will  confirm the features of the supersymmetric KKLT construction presented in  \cite{Bergshoeff:2015jxa}, and we will  describe the  positions of the O3-plane and the properties of the warped throat geometries for which the analysis in \cite{Bergshoeff:2015jxa} can be applied. This aspect of the work is very important for the eventual analysis of the possible values our parameter $M$ can take in string theory, so that the string landscape picture can be supported for our supersymmetric KKLT case.

Thus, the main purpose of this note is to describe explicit realizations of string theory anti-D-brane sectors with a spectrum corresponding to the nilpotent chiral multiplet with a fermionic goldstino which has no bosonic superpartners. The absence of scalars, which follows from the consistent orientifolding,  shows that these local models have improved stability properties. These local  models describe only the $\aD3$-brane at warped orientifold throat, but this should be regarded as a sector of a fully-fledged compact Calabi-Yau compactification. Such global models of string compactification can then achieve the realization of 
the phenomenological models of inflation and supersymmetry breaking in dS vacua proposed in  \cite{Kallosh:2014via,Carrasco:2015pla}, and other extensions thereof.

As mentioned, the case of a single $\aD3$-brane seems to enjoy improved stability properties. For instance, 
a special role of a single $ \aD3 $ in the KKLT uplifting was recently established using the effective field theory (EFT) methods in \cite{Michel:2014lva}, where it was argued that EFT description allows to use the brane actions beyond the probe approximation. For a single $\aD3$-brane in a flux threaded KS throat the back reaction is small and there is no instability at small string coupling, in accordance with earlier studies in \cite{Kachru:2002gs}.   
 It is moreover possible that beyond the most clear case of  one $\aD3$-brane, the comments in [20] hold for multiple antibranes if  $g_s N << 1$ (with N the number of anti-branes), in agreement with the suggested existence of a metastable state in this regime in the probe analysis of \cite{Kachru:2002gs}.

Meanwhile, 
during the last  years there were  papers studying KKLT construction with many $ \aD3 $ branes using a d=10 supergravity approximation and the feedback on it from many branes. There are  many examples of tachyonic instabilities and singularities,  see for example  \cite{Van}.
However, such an analysis is based on the classical 10d supergravity approach which requires small string coupling $g_s \ll 1$ and a large product $g_s  N\gg1$ where $N$ is the number of antibranes, hence they require $N \gg {1\over g_s}\gg 1$. It means that  these investigations reviewed are valid only for a large number of branes, and the approximation is not reliable for our case of a single  $\aD3$-brane.

For the supersymmetric  single $ \aD3 $-brane studied in  \cite{Bergshoeff:2015jxa} on the basis of $\kappa$-symmetry, there is no known non-Abelian generalization to the case of many branes. It is sufficient to remind that even the non-Abelian version of the bosonic Born-Infeld action is not available.
There is therefore a significant difference between the single and many brane cases: at our present level of knowledge  we can only state that the supersymmetric KKLT uplift in  \cite{Bergshoeff:2015jxa} is available for a single $ \aD3 $ brane.  

In this paper we will focus on a single $ \aD3 $, and exploit the microscopic string theory description to provide further insights into the stability of the system, and potential non-perturbative decay channels. In addition we will also comment on the mutiple $\aD3$-brane from the perspective of the string spectrum.

The rest of the presentation is organized as follows. Section 2 is devoted to general string theoretical realizations of $\aD3$-branes in orientifolded  flux backgrounds. The overall differences and similarities of the spectrum on D3- and $\aD3$-branes are emphasized. For ease of presentation, we assume the presence of a warped throat (induced by fluxes) and postpone their explicit construction to later sections. We describe the computation of the worldvolume spectrum on a single $\aD3$-brane stuck at an orientifold plane, and explore the result for O3- or O7-planes. We show that a combination of fluxes and orientifold projection leaves a single 4d massless fermion in the low-energy spectrum of the $\aD3$-brane, which can be identified as the goldstino. We explain the stringy origin of the non-linearity in goldstino, which is essential   for the spontaneously broken supersymmetry. The generalization to a higher number of anti-branes is briefly described. 

In section 3 we concentrate on the description of O3-planes at the tip of a throat. We show that, for the simple case of the KS throat,  orientifold action producing O7-planes are possible, but there are no consistent orientifolds producing O3-planes at the bottom of the throat. However, we describe the construction of more general warped throats admitting such O3-plane at their tip, and with the same warping behaviour as the KS case. We moreover provide an explicit example with an explicit holographic dual field theory, based on a deformed generalized conifold geometry. In section 4 we start a preliminary discussion on the expected couplings of the nilpotent superfield $X$ to K\"ahler moduli and chiral matter fields in the visible sector and discuss the expected value of the soft breaking terms. Finally in section 5 we present general discussions of the implications of these configurations for KKLT and related scenarios of moduli stabilization.

\section{String theory realization of the nilpotent goldstino}
\label{sec:string-minimal-goldstino}

In this section we provide the string theory construction of a local system of $\aD3$-branes on warped throats and show that the worldvolume spectrum contains only the goldstino of the broken supersymmetries, with no extra fields. This shows, with account of a non-linear goldstino coupling,  that the presence of the $\aD3$-brane breaks supersymmetry spontaneously. This also simplifies the description of systems including this kind of antibranes, by using the nilpotent goldstino multiplet to write supersymmetric actions \cite{Komargodski:2009rz}.

The construction is based on one $ \aD3 $-brane on top of an O3-plane at the bottom of a warped throat (or  in more generality, in the presence of imaginary self-dual (ISD) 3-form flux $G_3$). This is precisely the setup used in uplifting to de Sitter (in the KKLT or LARGE volume scenarios), and in the inflation models described in the introduction. 

In this section we only use general features of warped throats, like the presence of supersymmetric 3-form fluxes. Explicit examples will be discussed in section \ref{sec:throats}.

\subsection{D3- and $ \aD3$-branes on warped throats}
\label{sec:d3-spec}

We now describe the worldvolume spectrum on D3- and/or $\aD3$-branes on top of O3-planes. These computations are relatively standard, and we basically quote the results and their physical interpretation. 

\subsubsection{Open string spectra in 10d flat space}
\label{sec:flat10d}

As a warmup, consider a stack of $N$ D3-branes in flat 10d space. As is familiar \cite{Polchinski:1998rr}, the massless open string spectrum, classified according to representations of the $SO(3,1)$ 4d Lorentz group on the brane worldvolume, the $SO(6)\simeq SU(4)$ rotation group in the transverse dimensions, and the $U(N)$ gauge group, is shown in Table \ref{tab-dthree}.


\begin{table}[htb] 
\renewcommand{\arraystretch}{1.25}
\begin{center}
\begin{tabular}{cccc}
Field & $SO(3,1)$ & $SO(6)$ & $U(N)$  \\
Gauge boson & vector & ${\bf 1}$ & {\rm Adj} \\
Scalar & ${\bf 1}$ & {\bf 6} &  {\rm Adj} \\
Fermion & spinor & {\bf 4} &  {\rm Adj} \\
\end{tabular}
\end{center} 
\caption{ Spectrum on a stack of $N$ D3-branes in flat space}
\label{tab-dthree} 
\end{table}

It is the ${\cal N}=4$ $U(N)$ super Yang-Mills vector multiplet. The supersymmetry of the open string sector is related, by open-closed duality, to the BPS cancellation of NSNS and RR closed string exchange between parallel D3-branes, as follows. The one-loop open string partition function (annulus diagram) is given (up to a center of mass momentum factor) by
\beqa
Z_{\rm annulus}=\tr_{\cal H_{\rm open\; NS+R}} \big(q^{L_0}\big)
\eeqa
where $q=e^{-2\pi t}$ is the modular parameter and $L_0$ is  the open string Hamiltonian in the NS or R Hilbert space ${\cal H}_{\rm open}$.  The diagram can be transformed into a  tree level exchange of NSNS and RR closed string states between boundaries, i.e. D3-branes, with the structure
\beqa
Z_{\rm annulus}= \langle {\rm D3}|\, q'^{\,2L_0}\,
|{\rm D3}\rangle_{\rm NSNS}  + \langle {\rm D3}|\, q'^{\,2L_0}\,
|{\rm D3}\rangle_{\rm RR} 
\eeqa
 where now $L_0$ is the closed string Hamiltonian in the NSNS and RR sectors, the factor of 2 in the exponent accounts for left- and right-moving sectors, and $q'=e^{-2\pi t'}$ with $t'=1/t$.  

\medskip

If we instead consider a stack of $N$ $ \aD3 $-branes, we obtain precisely the same spectrum, as there is no way to distinguish D3- from $ \aD3 $-branes if they are isolated configurations  (see e.g. \cite{Ibanez:2012zz} for review). The only difference, since they preserve opposite set of supersymmetries, is that the fermions transform in the ${\bf \overline 4}$ of $SU(4)$, which amounts to a mere convention in the absence of extra ingredients. Of course, in real string compactifications there are many ingredients that distinguish them. In our case,  we are interested in branes located on warped throats supported by fluxes. We turn to consider their effect on the worldvolume spectrum. 

\bigskip

\subsubsection{Effects of fluxes in warped throats}

The masslessness of the above spectrum is in general modified in the presence of NSNS and RR 3-form fluxes $G_3$, such as those supporting the throat (or with more general fluxes introduced to stabilize compactification moduli) \cite{Dasgupta:1999ss,Giddings:2001yu}. 

We start by pointing out that such fluxes have no effect on the gauge group, so the gauge bosons remain massless. The effect of fluxes on the remaining massless sector was considered in \cite{Grana:2002nq,Camara:2003ku,Grana:2003ek} (see \cite{Bergshoeff:2015jxa} for a recent analysis of the action of the $ \aD3 $-brane action subject to orientifolding condition). We use a language from \cite{Camara:2003ku,Bergshoeff:2015jxa}. Consider first the fermions $\lambda$, which transform as a ${\bf 4}$ (or ${\bf \bar 4}$) under the $SO(6)$ rotation group. As shown by these references, the fermions pick up a mass term of the form
\beqa
G_3\, \lambda\lambda
\eeqa
The precise flux components providing mass for each of the four fermions follow from the $SO(6)$ selection rules. The flux density $G_3$ is a 3-index antisymmetric tensor, which decomposes into an imaginary self-dual (ISD) and imaginary antiself-dual (IASD) parts, transforming as a ${\bf 10}$ and ${\bf \overline {10}}$ of $SO(6)$, respectively. Therefore, the fermions on a D3-brane can couple (through ${\bf 4}\cdot {\bf 4}\cdot {\bf \overline{10}}$) to the IASD flux component, and remain massless in ISD fluxes. This is a consequence of the cancellation between contributions from the DBI and the CS actions, as checked explicity in the above references. 
On the other hand, we get the opposite result for $ \aD3 $-branes, whose fermions remain massless in the presence of IASD flux, but get masses (through ${\bf \bar 4}\cdot{\bf \bar 4}\cdot {\bf 10}$) in the presence of ISD fluxes.

To be more concrete about the fermion spectrum, we notice that the throat in \cite{Klebanov:2000hb} (as well as the more general throats in \cite{Franco:2004jz, Franco:2005fd}, c.f. section 3) are supported by supersymmetry preserving $(2,1)$ primitive 3-form flux. Decomposing $SO(6)\to SU(3)\times U(1)$ as befits 4d $\NN=1$ supersymmetry, we can decompose the fluxes as ${\bf 10}={\bf 6}+{\bf\overline 3}+{\bf 1}$ (and its conjugate decomposition for ${\bf \overline{10}}$), with the $(2,1)$ primitive flux corresponding to the ${\bf 6}$ (see e.g. Table 1 in \cite{Camara:2003ku}). Decomposing the fermions as ${\bf 4}={\bf 3}+{\bf 1}$ (for D3-branes) and ${\bf\overline 4}={\bf \overline 3}+{\bf 1}$ (for $ \aD3 $-branes), we see that the $(2,1)$ fluxes leaves the D3-brane fermions massless. However, it gives mass to the triplet of fermions on $ \aD3 $-branes, and leave precisely one massless 4d fermion. The intuitive explanation is that it corresponds to the goldstino of the 4d $\NN=1$ supersymmetry of the flux configuration, which is broken spontaneously by the introduction of the $\aD3$-brane. This model will play a crucial role in section \ref{sec:minimal}.

Finally, consider the scalars. We focus on the $n=1$ case, to have a minimal number of fields, but also to avoid polarization due to cubic terms in the scalar potential \cite{Kachru:2002gs}, which appear for $n>1$.

The scalars get masses quadratic in the flux density, in a way compatible, but not fully determined, by  $SO(6)$ quantum numbers. Hence we appeal to the results in \cite{Grana:2002nq,Camara:2003ku,Grana:2003ek}, focusing already on $(2,1)$ ISD flux. As already noted in \cite{Giddings:2001yu}, the result is that scalars on D3-branes get no mass, due to a BPS cancellation in the interaction of the D3-branes and the fluxes: the gravitational and electrostatic interactions cancel for D3-branes in supersymmetric flux, as they are mutually BPS \footnote{The cancellation occurs also (albeit accidentally) for the non-supersymmetric ISD  $(0,3)$ flux component, in agreement with the no-scale structure of the theory at tree level.}. On the other hand, for $ \aD3 $-branes, both effects add up, and the scalars get positive squared masses, stabilizing the $ \aD3 $-branes at the maximum of the warp factor (and flux density), the bottom of the throat \cite{Kachru:2002gs}.

Therefore, for one D3-brane in the presence of $(2,1)$ fluxes, the massless spectrum is given by a $U(1)$ gauge boson, six real scalars and four fermions. For one $ \aD3 $-brane in the presence of $(2,1)$ fluxes, the massless spectrum is given by a $U(1)$ gauge boson, and one 4d spinor. Below the flux scale (at which other states should be included in the effective action), this comes very close to the realization of a spectrum with just the goldstino.  It is really remarkable that this minimal goldstino structure  arises almost precisely in the configurations used for de Sitter uplifting.

In the following section we introduce a last ingredient to actually achieve a spectrum with just the goldstino.

\subsection{Anti-D3-brane with O3-plane and the nilpotent goldstino}
\label{sec:minimal}

In this section we introduce an O3-plane, and consider the worldvolume spectrum for a single D3- or a single 
$ \aD3 $-brane on top of the O3-plane\footnote{Initially we focus on the negatively charged O3-plane, denoted O3$^-$-plane, which corresponds to a Chan-Paton projection matrix $\gamma_\Omega={\bf 1}$; the O3$^+$-plane is discussed in section \ref{sec:more-general}.}. For a discussion of the embedding of these configurations in warped throats, see section \ref{sec:throats}.

\subsubsection{D3-branes with O3-plane}

Consider first the configurations in flat 10d space, and introduce one D3-brane on top of an O3-plane.  The introduction of the orientifold plane corresponds to modding out the theory by $\Omega'\equiv\Omega R (-1)^{F_L}$, where $F_L$ is spacetime left-moving fermion number, and $R$ is a geometric actiong flipping the sign of the dimensions transverse to the O3-plane, which is therefore the fixed point of this action. The single D3-brane has no orientifold image, and therefore it cannot move off the O3-plane; this will nicely dovetail with the worldvolume spectrum content computed below. Also, let us note that the configuration with a single D3-brane is possible only for the negatively charged  O3$^-$-plane; the choice of O3$^+$-plane can be discussed in section \ref{sec:more-general}.

The computation of the D3-brane worldvolume spectrum amounts to imposing an orientifold projection on the parent oriented D3-branes spectrum of section \ref{sec:d3-spec} (see e.g. \cite{Polchinski:1998rr,Ibanez:2012zz} for textbook reviews). Let  $\lambda{\cal O}|0\rangle$ denote a massless state of a parent D3-brane system, with $\lambda$ the Chan-Paton wavefunction, and ${\cal O}$ the NS or R oscillators. The action of the O3-plane is
\beqa
\lambda{\cal O}|0\rangle \;\rightarrow\; -(\gamma_{\Omega}\lambda^T\gamma_{\Omega}^{-1})\,{\cal O}|0\rangle
\label{projection-d3}
\eeqa
where $\gamma_\Omega$ is the action on Chan-Paton indices. This result is essentially identical to the (T-dual, in suitable toroidal compactifications) orientifold projection in type I theory. In particular, it is identical  to the projections in e.g. \cite{Pradisi:1988xd,Angelantonj:1996uy}, in eq.(3.11)-(3.13) of \cite{Gimon:1996rq}, or eq. (4.107) in \cite{Ibanez:2012zz}.

For a single D3-brane, Chan-Paton matrices reduce to complex numbers, and for an O3$^-$-plane we have $\gamma_\Omega=1$. All the massless states, both for NS and R states, are odd under the orientifold action, and therefore are projected out, so the D3-brane has no degrees of freedom at all. The physical interpretation is that, since the D3-brane is stuck at the O3-plane, there are no massless scalars in the spectrum; then, since the D3-brane preserves the same 16 supersymmetries as the O3-plane, the orientifold projection must remove the whole 4d $\NN=4$ vector multiplet, i.e. the gauge bosons and fermions as well.

\begin{figure}[htb]
\begin{center}
\includegraphics[scale=.5]{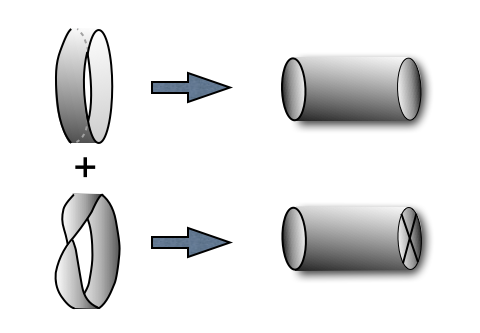} 
\caption{\small The one-loop open string annulus and Moebius strip diagrams turn into closed string channel diagrams describing tree level exchange of NSNS and RR states between two boundaries (branes or antibranes), or between one boundary and one crosscap (O3-plane).}
\label{fig:open-closed}
\end{center}
\end{figure}

The supersymmetry of the orientifold projection in the open string channel is related, by open-closed duality, to the cancellation of  NSNS and RR exchanges in the closed string channel. For the annulus diagram, this works as in the parent oriented theory; for the Moebius strip diagram, which is responsible for the orientifold projection, this corresponds to the BPS cancellation of the gravitational and 4-form interactions between the D3-brane and the O3-plane, see Figure \ref{fig:open-closed}.

More concretely, combining  the  open string annulus and Moebius strip amplitudes we have
\beqa
Z_{\rm annulus}+Z_{\rm moebius}=\tr_{\cal H_{\rm  open\; NS+R}} \big(q^{L_0}\,\big)+\tr_{\cal H_{\rm open\; NS+R}} \big(q^{L_0}\, \Omega'\big)=2\,
\tr_{\cal H_{\rm open\; NS+R}} \Big[q^{L_0}\, \frac{(1+\Omega')}{2}\Big]
\nonumber 
\eeqa
The operator $(1+\Omega')/2$ implements the projection onto invariant states. 

The transformation onto the closed string channel for the annulus  works as in section \ref{sec:flat10d}. For the Moebius strip, we have
\beqa
Z_{\rm Moebius}= \langle {\rm D3}| q'^{\,2L_0}
|{\rm O3}\rangle_{\rm NSNS} + \langle {\rm D3}| q'^{\,2L_0}
|{\rm O3}\rangle_{\rm RR}
\label{moebius-closed-d3}
\eeqa
where $q'=e^{-2\pi t'}$ with now $t'=1/(8t)$. For future convenience, we note that the Ramond sector Moebius strip amplitude translates into the RR exchange in the closed string channel.

\medskip

\subsubsection{Anti-D3-branes with O3-plane}

We now consider one $ \aD3 $-brane on top of the O3$^-$-plane. Again, the worldvolume spectrum is obtained by a simple orientifold action on the parent oriented spectrum in section \ref{sec:d3-spec}. This is very similar to the D3-brane case, except for the fact that the orientifold does not preserve the same supersymmetries as the $ \aD3 $-brane. As studied in \cite{Sugimoto:1999tx,Uranga:1999ib} (see also \cite{Antoniadis:1999xk,Aldazabal:1999jr},  and \cite{Ibanez:2012zz} for review), this manifests in an extra sign in the orientifold action on the open string Ramond sector. Namely we have
\beqa
& {\rm NS (bosons)}\quad &\quad \lambda{\cal O}|0\rangle \;\rightarrow\; -(\gamma_{\Omega}\lambda^T\gamma_{\Omega}^{-1})\,{\cal O}|0\rangle \nonumber\\
& {\rm R (fermions)}\quad &\quad \lambda{\cal O}|0\rangle \;\rightarrow\; +(\gamma_{\Omega}\lambda^T\gamma_{\Omega}^{-1})\,{\cal O}|0\rangle 
\label{projection-antid3}
\eeqa
The extra sign in the orientifold projection for Ramond states is easily derived from open-closed duality, as follows. The extra sign is in the Moebius strip diagram for open string Ramond states, which maps to an extra sign in the RR exchange between the crosscap and the boundary. This precisely matches the fact that $ \aD3 $-brane carry the same tension  as D3-branes, but opposite RR charge, hence the NSNS exchange is identical in both systems, but the RR exchange must have opposite signs. 

More concretely, the Moebius strip closed string channel amplitude (\ref{moebius-closed-d3}) turns into
\beqa
Z_{\rm Moebius}= \langle {\rm D3}| q'^{\,2L_0}
|{\rm O3}\rangle_{\rm NSNS} - \langle {\rm D3}| q'^{\,2L_0}
|{\rm O3}\rangle_{\rm RR}
\eeqa
And the extra sign propagates to the open string channel as
\beqa
&Z_{\rm annulus}+Z_{\rm moebius}&=\tr_{\cal H_{\rm  open\; NS}} \big(q^{L_0}\,\big)+\tr_{\cal H_{\rm  open\; R}} \big(q^{L_0}\,\big)+\tr_{\cal H_{\rm open\; NS}} \big(q^{L_0}\, \Omega'\big)-\tr_{\cal H_{\rm open\; R}} \big(q^{L_0}\, \Omega'\big)=\nonumber \\
&& =2 \tr_{\cal H_{\rm open\; NS}} \Big[q^{L_0}\, \frac{(1+\Omega')}{2}\Big]+ 2 \tr_{\cal H_{\rm open\; R}} \Big[q^{L_0}\, \frac{(1-\Omega')}{2}\Big]
\nonumber 
\eeqa
showing the different orientifold projection in the NS and R sectors.

Imposing this orientifold projection on the $ \aD3 $-brane spectrum, massless bosonic states are odd and therefore removed by the orientifold projection, exactly as in the D3-brane case.  On the other hand, due to the sign flip, the massless fermionic states are orientifold-even and remain in the spectrum. Thus a single $ \aD3 $-brane on top of an O3-plane has 4 fermions as its only worldvolume degrees of freedom.

The result of the orientifold projection has a nice physical interpretation. As in the D3-brane case, the scalars are absent because  the single D3-brane is stuck and cannot move off the O3-plane. In this case, we cannot exploit supersymmetry to argue for the disappearance of the gauge bosons, but it can be understood directly, as follows. Recall that the worldvolume gauge fields on D-branes are intimately related to the NSNS 2-form field, by the gauge symmetry
\beqa
B_{\mu\nu} \to B_{\mu\nu}+\partial_{[\mu} \Lambda_{\nu]} \quad ;\quad A_\mu\to A_\mu+\Lambda_\mu
\eeqa
The familiar fact that the NSNS 2-form is projected out by orientifolds is thus correlated with the removal of the worldvolume gauge field on D-branes, in our case on D3-branes (in agreement with supersymmetry, as discussed above) or on $ \aD3 $-branes, as in our case of present interest.

Finally, the four massless fermions that remain in the spectrum are the goldstinos of the sixteen supersymmetries preserved by the O3-plane, which are all spontaneously broken by the introduction of the $ \aD3 $-brane. 

\medskip

As in the last part of section \ref{sec:d3-spec}, we are actually interested in brane systems in the presence of $(2,1)$ supersymmetric fluxes. The effect of these fluxes is simply obtained by considering the fluxes in the parent theory, and truncating the spectrum by the orientifold projection. More explicitly, for a single $ \aD3 $-brane in the presence of $(2,1)$ fluxes, the $SU(3)$ triplet of fermions is made massive by the fluxes, while exactly one fermion remain massless. 

In conclusion, for a single $ \aD3 $-brane on top of an O3-plane, in the presence of  (2,1) fluxes (such as those in warped throats), the light spectrum below the flux scale is given by exactly one massless fermion. This is interpreted as the goldstino of the 4d $\NN=1$ supersymmetry preserved by the O3-plane and the fluxes, broken spontaneously by the presence of the $ \aD3 $-brane. Therefore this setup provides an explicit string theory realization of a spectrum given exactly by the nilpotent chiral multiplet. The fact that supersymmetry with only fermions in the spectrum is broken spontaneously requires an additional confirmation: we will argue below that the non-linear terms in fermions are present in the action of the  $ \aD3 $-brane.

A natural question at this point is whether the spontaneously broken supersymmetries are restored at higher energies. In a sense, the answer in the full string theory lies in the earlier discussions of open-closed duality. The open string diagrams sensitive to the supersymmetry breaking, like the Moebius strip, have a UV behaviour which is controlled by the closed string channel, which is supersymmetric. Clearly, this regime is beyond the reach of effective field theory, which therefore contains only the degrees of freedom corresponding to the goldstino.

Besides the above described vanilla-type  setup, it is also possible to achieve this spectrum from $\aD3$-branes on other orientifold planes, like O7-planes. These setups will be described in section \ref{sec:o7plane}

\subsection{Non-linear goldstino interactions}

From the spectrum of the configuration of a single $ \aD3 $-brane on top of an O3-plane we have inferred that only a single Majorana fermion is present. To argue that supersymmetry is broken spontaneously as in Volkov-Akulov (VA) theory \cite{Volkov:1973ix}, we have to explain the  origin of non-linear terms beyond the kinetic terms for the fermions.  For this purpose we refer to a property of the parent theory of a single $ \aD3 $-brane before orientifolding. 

Recall how the non-linear interactions of the D-brane gauge bosons are nicely resummed in a  Born-Infeld expression $\sqrt {(1+ (\alpha' F)^2)}$,  as can be shown from the string partition function in the presence of a single magnetic field component $F$, see e.g. \cite{Tseytlin:1999dj}. Indeed, the $F^2+ \alpha^{'2} F^4$  terms in BI action are in precise agreement with the ones derived directly from (super)string 4-point amplitude \cite{Gross:1986iv}. This motivated the suggestion in \cite{Tseytlin:1999dj} to use  4d ${\cal N}=2$ and ${\cal N}=1$ superfields to obtain the non-linear fermionic partners to the non-linear Born-Infeld terms $F^{2n}$. Namely the supersymmetrization of the Born-Infeld terms e.g. of the form $F^4$ gives the result one would obtain from direct computation of the  (super)string 4-point amplitude for fermions. 

An advanced version of this kind of supersymmetrization was presented in \cite{Bergshoeff:2013pia}, allowing for the computation of the local terms in the D-brane action producing higher order non-linear terms both in Maxwell field as well as in spinor fields.
 For example, the action of D3 or $ \aD3 $-brane in a gauge where WZ term is absent and only DBI term is present has the following form \cite{Bergshoeff:2013pia}
\begin{equation}
\label{4actionBI} S =  -\frac{1}{\alpha^2} \int \rmd^{4} x\,
\sqrt{- \det (G_{\mu\nu} + \alpha {\cal F}_{\mu\nu})}  \,, \qquad \mu=0,1,2,3\,,
\end{equation}
where
\begin{eqnarray}
G_{\mu\nu} &=& \eta_{m n} \Pi_\mu^{m}
\Pi_\nu^{n}= \eta_{m' n'} \Pi_\mu^{m'}
\Pi_\nu^{n'}+\delta_{IJ} \Pi_\mu^I\Pi_\nu^J \,, \qquad  m'=0,1,2,3 \,, \qquad I=1,...,6.\nonumber \\[.2truecm]
\Pi_\mu^{m'} &=& \delta^{m'}_{\mu}
-\alpha^2\bar\lambda \Gamma^{m'}
\partial_\mu \lambda \ , \qquad \Pi_\mu^I = \partial_\mu\phi^I
-\alpha^2\bar\lambda \Gamma^I
\partial_\mu \lambda\,, \qquad {\cal F}_{\mu\nu} \equiv F_{\mu\nu} - b_{\mu\nu} \,,\nonumber\\
 b_{\mu\nu} &=&2\alpha\bar{\lambda}\Gamma_{[\mu }\partial_{\nu]}\lambda
-2\alpha\bar{\lambda}\Gamma_{I}\partial_{[\mu}\lambda\partial_{\nu]}
\phi^I =-2\alpha\bar{\lambda}\Gamma_{m'}\partial_{[\mu}\lambda\, \Pi_{\nu]}^{m'}
-2\alpha\bar{\lambda}\Gamma_{I}\partial_{[\mu}\lambda \, \Pi_{\nu]}
^I  \, .
\end{eqnarray}
This action has a maximal number of  supersymmetries, namely 16
$\epsilon$-supersymmetries correspond to a deformation of the original
16 supersymmetries of the ${\cal N}=4$, $d=4$ Maxwell multiplet, and 16 
$\zeta$-supersymmetries correspond to a Volkov-Akulov-type supersymmetries.
For example, the $\zeta$ transformations on the brane acting on the spinor $\lambda$ are 
\be\label{eq:lambda}
\delta_\zeta \lambda =  \zeta +  \bar \lambda \gamma^a \zeta \partial_a \lambda\,,
\ee

If instead of the regular $\kappa$-gauge symmetry fixing one imposes the orientifolding condition as explained in \cite{Bergshoeff:2015jxa}, the action of the $ \aD3 $-brane reduces to a VA action without vectors and scalars \footnote{A more detailed study of the fermionic terms on the $\aD3$-brane reveals
 that the VA form for the $\aD3$-brane action may persist
 even when all orders of fermionic interactions are taken into account
\cite{KT}.} . Namely, a direct computation of superstring amplitudes in a situation with only fermions, and  no vectors or scalars, would reproduce  the non-linear terms in the  $ \aD3 $-brane of the form \cite{Bergshoeff:2015jxa}
\be
S_{VA}= - M^2  \int d^4 \sigma \det E  = - M^2  \int E^0\wedge E^1 \wedge E^2  \wedge E^3
\,, \qquad  E^a = \delta^a_\mu d\sigma^\mu + \bar \lambda \gamma^a d\lambda\,,
\label{VA}\ee
This action is invariant under the  non-linear ${\cal N}=1$ supersymmetry transformation \rf{eq:lambda}.
The action can be also  given in the following form (see \cite{Komargodski:2009rz}, where it was derived for a nilpotent superfield, $X^2(x, \theta)=0$)
\be
{\cal L}_{KS}= - M^2 +\rmi \partial_a \bar \psi \bar \sigma^a \psi + {1\over 4 M^2} \bar \psi^2 \partial^2 \psi^2 - {1\over 16 M^{6}} \psi^2 \bar \psi^2 \partial^2 \psi^2 \partial^2 \bar \psi^2 \,,
\label{VA1}\ee
with $\bar{\sigma}^a = (-\mathbb{1}, -\sigma_n)$. As it is shown explicitly in \cite{Kuzenko:2010ef}, the above action agrees with the original VA action \eqref{VA} after a spinorial field redefinition  $\psi=M^2 \lambda$ plus terms non-linear in fermions.

\subsection{Additional $\aD3$-branes and other variants}
\label{sec:more-general}

In this section we briefly consider how the earlier discussion changes for other systems of $\aD3$-branes and O3-planes. These variants were considered in \cite{Uranga:1999ib}, see table below\footnote{This corrects a typo in  \cite{Uranga:1999ib}.}, and are included here for completeness. The bottom-line is that for O3-planes, the only setup realizing just the goldstino multiplet is that considered in the previous sections.

\begin{center}
\begin{tabular}{|l||c|c|c|}
\hline
             & $SO(1,3)$ & $SO(6)$ & O3$^-$: $SO(N)\, $ ; O3$^+$: $USp(N)\, $ \\
\hline\hline
Gauge bosons & vector & singlet & $N(N-1)/2$ ; $N(N+1)/2$\\
\hline
Scalars & singlet & vector &  $N(N-1)/2$ ; $N(N+1)/2$ \\
\hline
Fermions & spinor & spinor &  $N(N+1)/2$ ; $N(N-1)/2$ \\
\hline
\end{tabular}
\end{center}

\subsubsection{Additional D3- or $\aD3$-branes}

Consider a stack of $N$ D3- or $\aD3$-branes on top of an O3$^-$-plane in flat 10d space. The massless spectra are obtained by applying the orientifold projection conditions (\ref{projection-d3}) or  (\ref{projection-antid3}), but with an enlarged $N\times N$ Chan-Paton matrix. 

For $N$ D3-branes, all NS and R massless states project onto antisymmetric Chan-Paton matrices. The massless spectrum thus corresponds to $SO(N)$ gauge bosons, six scalars and four fermions, all in the antisymmetric (i.e adjoint) representation. This spectrum fills out the 4d $\NN=4$ $SO(N)$ vector multiplet, since D3-branes preserve the same supersymmetry as the O3-plane.

For $N$ $\aD3$-branes, the NS massless states project down to the antisymmetric representation, but due to the extra sign for Ramond states project down onto the symmetric representation. Therefore we obtain $SO(N)$ gauge bosons, six scalars in the antisymmetric representation, and four fermions in the symmetric representation. 

Note that for $N=1$ the antisymmetric representations are empty, and we recover the spectrum of the previous section, with a purely fermionic massless spectrum, reproducing the goldstino of the spontaneously broken supersymmetries. Note also that the two-index symmetric tensor representation of $SO(N)$ is reducible, and splits into a singlet (the trace) and a traceless symmetric tensor. The singlet of fermions also corresponds to a goldstino, for general $N$, and could be described in terms of a nilpotent chiral multiplet. However, the presence of the extra fields (with non-matched fermions and bosons) requires the introduction of additional constrained superfields, as introduced in \cite{Komargodski:2009rz}. Since these systems do not reproduce the minimal goldstino setup, we refrain from discussing them further.

Moreover, as reviewed in the introduction, multiple $\aD3$-brane systems have additional difficulties. Namely, they do not have a known realization of the local $\kappa$-symmetry. Also, it is easy to show that they suffer a repulsive instability, additional pairs of $\aD3$-branes located on top of the O3-plane are dynamically expelled off, as follows. Since both the tension and 4-form charge of the O3$^-$-plane are negative, both the gravitational and RR exchange interaction between the O3-plane and the $\aD3$-brane pair are repulsive. The above closed string channel description has an open string channel description as well: the one-loop quantum correction to the masses of scalars that appear in multiple brane systems renders them tachyonic and thus unstable. 

This analysis suggests that the configuration with one $\aD3$-brane on top of the O3-plane is natural (as others dynamically flow to it, by pairwise expulsion of additional $\aD3$-branes), and constitutes a stable configuration (as it has no scalars and thus no further tachyons). Note that for the stuck antibrane, the repulsive interaction is just a constant contribution to the vacuum energy, i.e. a (finite) quantum correction to the cosmological constant.

These observations actually emphasize the simplicity and elegance of the single $\aD3$-brane on a O3-plane, where the $\kappa$-symmetry has been studied in \cite{Bergshoeff:2015jxa}, and the absence of scalars ensures the stability of the systems agains tachyons.  This nicely dovetails the fact that the system with a single $\aD3$-brane is not well-described as a supergravity solution, and therefore can evade the problems found there.

\subsubsection{Using O3$^+$-planes}

For completeness we comment of the system of D3- and $\aD3$-branes on top of a positively charged, O3$^+$-plane. The definition of a string theory configuration in the presence of an O3$^+$-plane is identical to the O3$^-$-plane case, except that the quantum amplitude of a worldsheet with $n_c$ crosscaps has an extra sign  $(-1)^{n_c}$ (see \cite{Ibanez:2012zz} for a textbook review). This is easily understood by considering the tree-level closed string exchange between one crosscap and one boundary: the overall sign change due to the crosscap reflects the fact that an O3$^+$-plane has opposite tension and charge as compared with the O3$^-$-plane. 

The extra sign flips the orientifold projection in the open string (both NS and R) sector. The resulting massless spectra for $N$ D3- or $\aD3$-branes on top of an O3$^+$-plane are as follows. For $N$ D3-branes, we get gauge bosons of $USp(N)$, six scalars and four fermions, all in the two-index symmetric  (i.e. adjoint) representation. This fills out the 4d $\NN=4$ $USp(N)$ vector multiplet, in agreement with the mutual supersymmetry of D3-branes and O3-planes. Note that $n$ is forced to be even for consistency, therefore the minimal non-trivial case of this kind corresponds to $N=2$.

For $N$ $\aD3$-branes we get $USp(N)$ gauge bosons, six real scalars in the symmetric representation, and four fermions in the antisymmetric representation. Considering the minimal case of $N=2$, the latter are gauge singlets, and correspond to the goldstinos of the spontaneously broken supersymmetries (for higher $n$, the antisymmetric representation is reducible and also contains gauge singlets corresponding to the goldstinos). However, these goldstinos are accompanied by non-Abelian gauge bosons and charged scalars, so the spectrum does not correspond to purely the nilpotent chiral multiplet, and may suffer from the difficulties of the $N>1$ case mentioned in the previous section.

As described in \cite{Uranga:1999ib, Sugimoto:2012rt}, the system of two $\aD3$-branes on top of an O3$^+$-plane describes the S-dual, infinite coupling regime of the single $\aD3$-brane on top of the O3$^-$-plane. The presence of additional scalars might seem bothersome, but they are easily shown to actually be very massive. In the open string channel, this follows from the one-loop quantum correction tot he scalar mass. In the closed string channel, it follows from the attractive intraction between the O3$^+$-plane and the two $\aD3$-branes. However, note that even if scalars are absent, the configuration still contains $SU(2)$ gauge bosons and therefore is not a realization of a configuration with just the goldstino multiplet.

In any event, these issues are directly absent in the weak coupling regime of the single $\aD3$-brane stuck at the O3$^-$-plane, at which we assume the dilaton is stabilized by bulk fluxes.

\subsection{The nilpotent goldstino from $\aD3$-branes with an O7-plane}
\label{sec:o7plane}

In this section we consider the configuration of single $ \aD3 $-brane in the presence of other orientifold planes, like O7-planes. As we will see, it can also produce the minimal spectrum of  just the goldstino at the massless level, albeit with additional massive scalars at the flux scale. 

We start in flat 10d space, split as 4d Minkowski space and three complex coordinates $z,w_1,w_2$. We introduce an O7-plane defined as the fixed point set under an orientifold action $\Omega {\cal R}(-1)^{F_L}$, with $F_L$ left-moving spacetime fermion number, and a geometric action ${\cal R}: z\to -z$ \footnote{For the initiated, note that there are two variants of O7-planes, which differ on their orientifold projection on the open string sector. We focus on the O7$^+$-plane, for reasons that will soon become clear.}. 

Consider a single D3-brane stuck at $z=0$ i.e. on top of the O7-plane. The D3-brane preserves 8 of the 16 supercharges of the O7-plane, so the spectrum fills out 4d $\NN=2$ multiplets. The computation is similar to that of D5-branes in the presence of an O9-plane (i.e. in type I theory), see \cite{Witten:1995gx}. The projections are as follows
\beqa
& {\rm Gauge \; bosons,\; scalar}\; z\quad &\quad \lambda{\cal O}|0\rangle \;\rightarrow\; -(\gamma_{\Omega}\lambda^T\gamma_{\Omega}^{-1})\,{\cal O}|0\rangle \nonumber\\
& {\rm Scalars}\; w_1,w_2\quad &\quad \lambda{\cal O}|0\rangle \;\rightarrow\; + (\gamma_{\Omega}\lambda^T\gamma_{\Omega}^{-1})\,{\cal O}|0\rangle \nonumber\\
& {\rm Two \; fermions}\; \lambda, \psi\quad &\quad \lambda{\cal O}|0\rangle \;\rightarrow\; -(\gamma_{\Omega}\lambda^T\gamma_{\Omega}^{-1})\,{\cal O}|0\rangle \nonumber \\
&{\rm Two \; fermions}\; \psi_1, \psi_2\quad &\quad \lambda{\cal O}|0\rangle \;\rightarrow\;  +(\gamma_{\Omega}\lambda^T\gamma_{\Omega}^{-1})\,{\cal O}|0\rangle 
\label{projection-antid3}
\eeqa
with $\gamma_{\Omega}=1$ for an O7$^+$-plane.
The orientifold removes the gauge boson, the scalar $z$, and the fermions $\lambda$, $\psi$. This can be understood because the D3-brane is stuck at $z=0$ so the corresponding scalar (and by supersymmetry the whole $\NN=2$ multiplet hosting it) must be projected out. The massless spectrum is given by the orientifold-even states, which correspond to two fermions $\psi_1$, $\psi_2$, and two scalars $w_1$, $w_2$. The corresponding 4d $\NN=2$ hypermultiplets is associated to the fact that the D3-brane can slide along the O7-plane in the directions $w_1$, $w_2$. As usual, this spectrum is not modified by the presence of ISD $(2,1)$ fluxes due to the relative BPS nature of the D3-branes and the fluxes.

\medskip

Consider now introducing a single $\aD3$-brane instead. By the same argument as in section \ref{sec:minimal}, the spectrum is obtained by imposing the orientifold projections, with an extra sign on the Ramond sector. Namely we have
\beqa
& {\rm gauge \; bosons,\; scalar}\; z\quad &\quad \lambda{\cal O}|0\rangle \;\rightarrow\; -(\gamma_{\Omega}\lambda^T\gamma_{\Omega}^{-1})\,{\cal O}|0\rangle \nonumber\\
& {\rm Scalars}\; w_1,w_2\quad &\quad \lambda{\cal O}|0\rangle \;\rightarrow\; + (\gamma_{\Omega}\lambda^T\gamma_{\Omega}^{-1})\,{\cal O}|0\rangle \nonumber\\
& {\rm Two \; fermions}\; \lambda, \psi\quad &\quad \lambda{\cal O}|0\rangle \;\rightarrow\; +(\gamma_{\Omega}\lambda^T\gamma_{\Omega}^{-1})\,{\cal O}|0\rangle \nonumber \\
&{\rm Two \; fermions}\; \psi_1, \psi_2\quad &\quad \lambda{\cal O}|0\rangle \;\rightarrow\;  -(\gamma_{\Omega}\lambda^T\gamma_{\Omega}^{-1})\,{\cal O}|0\rangle 
\label{projection-antid3}
\eeqa
The orientifold removes the gauge boson, the scalar $z$, and the fermions $\psi_1$, $\psi_2$. There survive four real scalars (the complex $w_1$, $w_2$) and two fermions $\lambda$, $\psi$.  These correspond to a 4d $\NN=2$ supersymmetry preserved by the $\aD3$-brane and the O7-plane (given by the 8 supersymmetries preserved by the O7-plane but broken by the D3-brane in the earlier computation). Hence, this $\NN=2$ is not preserved by other ingredients in string compactifications, and in particular is not compatible with the the (2,1) fluxes on warped throats. In fact, it is easy to check (using the $SO(6)$ quantum numbers, and their decomposition under the $SO(4)\times SO(2)$ preserved by the O7-plane) that the (2,1) fluxes give mass to the fermion $\psi$, and to the scalars $w_1$,$w_2$. The latter describes the fact that the $\aD3$-brane will be pinned to the point of maximal warp factor. 

In this situation, the massless spectrum contains only one fermion, which corresponds to the goldstino of the $\NN=1$ supersymmetry preserved by the fluxes (and the O7-plane) and are broken spontaneously by the $\aD3$-brane. Notice however the important difference with respect to section \ref{sec:minimal}, that in this case the scalars are not directly projected out, but remain in the massive spectrum, with mass fixed by the scale at the bottom of the throat. 

On the other hand, an advantage of the O7-plane setup is that it is very easy to construct warped throats admitting this kind of orientifold involution. This is possible for O3-planes, although it requires some heavier machinery. We turn to this discussion in the next section.

\section{Orientifolding warped throats}
\label{sec:throats}

The basic ingredient to realize the nilpotent goldstino spectrum are D3-branes and orientifold planes at the bottom of warped throats. In earlier sections, such throats were asummed to exist and to have the standard properties associated to the prototypical KS throat \cite{Klebanov:2000hb}, mainly the existence of $(2,1)$ 3-form fluxes. In this section we discuss the realization of throats of this kind with orientifold planes (of various kinds) at their bottom. 

\subsection{Preliminary remarks}

Let us consider the question of locating an O3-plane at the bottom of warped throats. One may naively think that this requires a substantial tuning. However, this is not the case, because the location of O3-planes is not determined by a choice of free parameters, but rather by the determination of fixed points under the geometric part ${\cal R}$ of the orientifold action $\Omega {\cal R}(-1)^{F_L}$. Locally ${\cal R}$ acts by flipping three local coordinates $(z_1,z_2,z_3)\to (-z_1,-z_2,-z_3)$. Actually the right description on a curved CY is to require that ${\cal R}$ flips the sign of the holomorphic 3-form (the earlier explicit action satisfies this, since locally the holomorphic 3-form is $dz_1dz_2dz_3$).

On the other hand, it is certainly true that not all warped throats admit an O3-plane symmetry. 
For instance, the well-established Klebanov-Strassler throat is based on a warped version of the deformed conifold geometry, which in complex variables is described by
\beqa
xy-zw=\epsilon
\label{defcon}
\eeqa
where $\epsilon$ is the size of the ${\bf S}^3$, fixed in terms of the flux data. One may be tempted to consider the ${\bf Z}_2$ action flipping all coordinates in this geometry,  
\beqa
(x,y,z,w) \to (-x,-y,-z,-w)
\eeqa
This is a symmetry (\ref{defcon}) for arbitrary $\epsilon$, and would seem to provide an unoriented version of the deformed conifold throat. However it does not introduce O3-planes, since the action has no fixed points, since the origin $x=y=z=w=0$ is not part of the CY defined by (\ref{defcon}). Moreover, the action is not acting properly on the holomorphic 3-form $dx\, dy\, dz/z$.

This is in fact a general feature of the KS throat: the deformed conifold does not have any holomorphic $\IZ_2$ symmetries with isolated fixed points.   Therefore the usual warped deformed conifold geometry does not admit the introduction of an O3-plane.
This is important to appreciate that the introduction of O3-planes is not a choice that we can control at will.

On the other hand, there exist orientifold actions which define O7-planes on the deformed conifold \cite{Park:1999ep}. For instance
\beqa
(x,y,z,w)\to (-x,-y,z,w)
\eeqa
This defines an O7-plane located at $x=y=0$, namely at $zw=-\epsilon$. This can be used to construct a string realization of the minimal goldstino multiplet as described in section \ref{sec:o7plane}, which is easily embedded in the KS throat. Namely, the O7-plane stretches in the radial direction getting down to the bottom of the throat, where the $\aD3$-brane sits.

\medskip

One should not conclude that O3-planes are too hard to be realized in warped throats. In the next section we address this question more systematically, and easily construct very explicit examples of throats (with explicit holographic dual quiver field theories) with O3-planes at their bottom.

\subsection{Explicit throats with O3-planes}

The reader not interested in the details, but just in one working example, can jump to the discussion around (\ref{orient-ecua-z2w3}).

\subsubsection{The conifold revisited}

In order to present the general analysis, we review the KS throat emphasizing those aspects important for  generalizations. 

\begin{figure}[htb]
\begin{center}
\includegraphics[scale=.5]{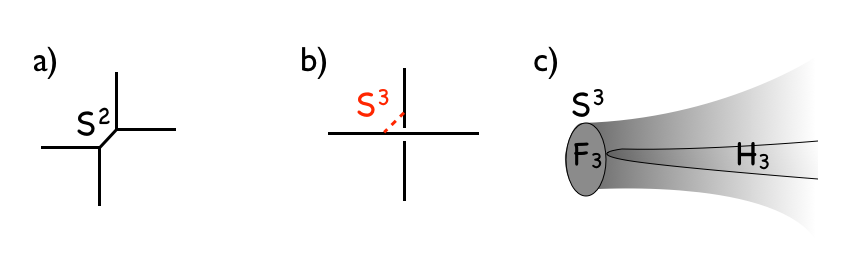} 
\caption{\small a) The web diagram for the resolved conifold; the finite segment corresponds to the 2-sphere. b) The splitting of the diagram into sub-webs in equilibrium describes the complex deformation; the dashed segment measuring the sub-web separation represents the 3-sphere. c) The 3-sphere in the complex deformation can support fluxes which lead to the KS warped throat.}
\label{fig:conifold}
\end{center}
\end{figure}

The KS throat is based on a warped version of the deformed conifold geometry. The latter can be obtained by a geometric transition from the resolved conifold. The resolved conifold is a toric geometry, i.e. it can be described as a series of circles fibered over a base, with loci where particular circle fibers (or combinations thereof) degenerate (see e.g. \cite{Iqbal:2001ye} for a simple introduction to toric geometry). The geometry is encoded in a so-called web diagram, which is a figure in a 2-plane (representing the base of the fibration), composed of lines and segments with a $(p,q)$ label defining their slope, and joining at vertices with an `equilibrium' sum rule: the $(p,q)$ labels of all incoming legs must add up to zero. The (p,q) label of each line representing that a $(p,q)$ linear combination of circle fibers degenerates over the locus of the base. The web diagram of the resolved conifold is shown in Figure \ref{fig:conifold}a.

The geometric transition of shrinking the 2-cycle and deforming the geometry by growing a 3-sphere corresponds to shrinking the $\IS^2$ finite segment in Figure \ref{fig:conifold}a, and separating the diagram into two sub-webs in equilibrium, the two indepent lines in Figure \ref{fig:conifold}b. The resulting geometry can support 3-form fluxes leading to the KS warped throat, shown in Figure \ref{fig:conifold}c.

The throat has a field theory dual, given by the quiver gauge theory of $N$ D3-branes in the resolved conifold geometry, in the presence of $M$ fractional  D5-branes wrapped on the $\IS^2$. In the gravity dual,  $M$ is the RR 3-form flux on the $\IS^3$, and $N$ is the total charge carried by the NSNS and RR fluxes, taking the throat to be cutoff at some radial distance to render the NSNS flux finite. The 4d $\NN=1$ theory, determined in \cite{Klebanov:1998hh} (see also \cite{Morrison:1998cs,Uranga:1998vf,Dasgupta:1998su}), has two gauge factors $SU(N)\times SU(N+M)$ with chiral multiplets $A_i$, $i=1,2$ in the $(\fund,\antifund)$ and $B_i$ in the $(\antifund,\fund)$, and a superpotential $W=\epsilon_{ij}\epsilon_{kl}\tr( A_iB_k A_j B_l)$. The warped throat is dual to the RG flow of this gauge theory, in which the gauge factors, in an alternating fashion, attain strong coupling and suffer a Seiberg duality, reproducing a periodic pattern known as duality cascade. The number of fractional branes is preserved, but the number of D3-branes decreases logarithmically along the RG flow. The infrared end of the RG flow arises when the D3-branes disappear, and the strong gauge dynamics on the remaining $M$ fractional branes confine and generate a dynamical scale, exponentially suppressed with respect to the UV cutoff, and whose gravity dual is the $\IS^3$ size, fixed by $\epsilon$ in (\ref{defcon}).

\begin{figure}[htb]
\begin{center}
\includegraphics[scale=.5]{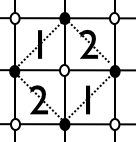} 
\caption{\small Dimer diagram for the theory of D-branes at a conifold. The dashed line is the unit cell in the periodic array.}
\label{fig:conifold-dimer}
\end{center}
\end{figure}

The gauge theories for branes at toric singularities are efficiently encoded in dimer diagrams \cite{Franco:2005rj}. They are (bipartite) graphs tiling an auxiliary $\IT^2$ (equivalently, doubly periodic two-dimensional graphs), with faces representing gauge factors, edges representing chiral multiplets in bi-fundamental representations of the adjacent gauge factors (with orientation given by moving e.g. clockwise around black nodes, and counterclockwise around white nodes), and nodes representing superpotential couplings of the chiral multiplets associated to the edges of the node (and sign determined by the node color). The diagram for the conifold is shown in Figure \ref{fig:conifold-dimer}. The web diagram corresponding to the toric geometry underlying a given dimer gauge theory is easily obtained by simple combinatorial tools, whose discussion we skip, directing the interested reader to the references. 

\subsubsection{More general throats and an explicit example}

In order to construct more general throats, we must consider more general toric singularities, which are easily engineered using the web diagrams. The quiver gauge theories associated to D-branes at such singularities can systematically studied using dimer diagram tecniques. However, not all such geometries admit complex deformations, and therefore not all can be used to define warped throats with smooth infrared ends. The criterion for the existence of complex deformations, and its dual field theory interpretation, were described in \cite{Franco:2007ii}. The result is that complex deformations correspond to splitting the web diagram into sub-webs in equilibrium. The field theory dual is described in terms of confinement on a set of gauge factors associated to certain fractional branes, and was systematically studied in terms of dimer diagrams in \cite{Franco:2007ii,GarciaEtxebarria:2006aq}. 

The geometries admitting complex deformations can be used to build throats, which are supported by (2,1) 3-form fluxes, with RR fluxes on the 3-cycles at the bottom of the throat, and NSNS fluxes on their dual 3-cycles. The field theory duals of these throats correspond to duality cascades triggered by the fractional branes dual to the RR 3-form fluxes, which lead to a reduction of the effective number of D3-branes as one runs to the infrared, and which ends in a process of confinement at a dynamical scale dual to the size of the infrared 3-cycles. The explicit construction of metrics for these throats depends on the ability to write metrics for the corresponding (deformed) cones, see \cite{Franco:2004jz} for some examples of the KT-like solutions for cones over del Pezzo surfaces. However, the main properties of the throats, like the existence of (2,1) fluxes, and the scaling of the warp factor with the flux quanta, can be established even with no information about the underlying metric. These properties are easily encoded in the existence of a supersymmetric RG flow representing a duality cascade.

\begin{figure}[htb]
\begin{center}
\includegraphics[scale=.5]{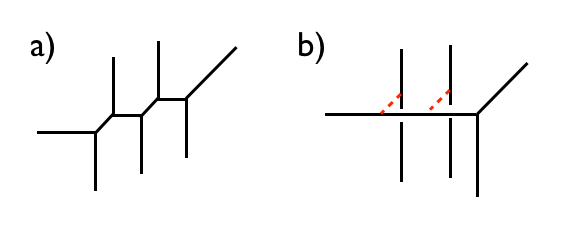} 
\caption{\small a) Web diagram for the geometry $xy=z^3w^2$ in the resolved phase; the exponents of $z,w$ in the defining equation are related to the numbers of parallel vertical external legs. b) Diagram for the deformed geometry, with two independent 3-spheres, shown as dashed lines.}
\label{fig:web-z2w3}
\end{center}
\end{figure}

As example, useful in later discussions,  consider the geometry $xy=z^3w^2$, whose web diagram in the resolved and deformed phases is shown in Figure \ref{fig:web-z2w3}. The resolved phase has two independent 3-spheres, which can support two different quanta of RR 3-form flux.

\begin{figure}[htb]
\begin{center}
\includegraphics[scale=.5]{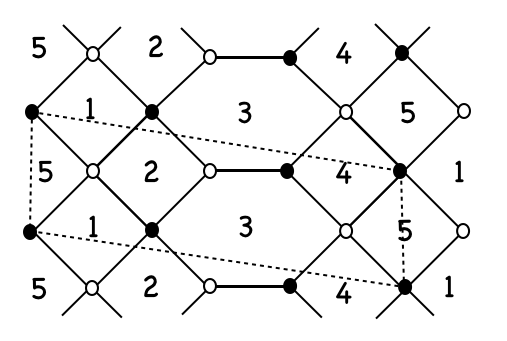} 
\caption{\small Dimer diagram describing the gauge theory on D3-branes at the  geometry $xy=z^3w^2$.}
\label{fig:dimer-z2w3}
\end{center}
\end{figure}

After the deformation, the geometry reads\footnote{A simple derivation is as follows. Consider the modified geometry $xy=z^3w^3$, which we can rewrite as $xy=t^3$, $t=zw$. The first equation is a $\IC^2/\IZ_3$ singularity, which can be deformed by splitting the zeroes of $t^3$, namely we end up with $xy=t(t+\epsilon_1)(t+\epsilon_2)$, $t=zw$, equivalently $xy=zw(zw+\epsilon_1)(zw+\epsilon_2)$. The deformation of $xy=z^3w^2$ is basically the same, by simply removing one power of $w$ (i.e. performing a partial blow-up).}
\beqa
xy =z (zw+\epsilon_1)(zw+\epsilon_2)
\label{ecuaz2w3}
\eeqa
The field theory associated to these singularities (first considered in  \cite{Uranga:1998vf}) has a dimer diagram shown in Figure \ref{fig:dimer-z2w3}. The fractional branes responsible for the duality cascade and complex deformation correspond to faces 2 and 4.

\begin{figure}[htb]
\begin{center}
\includegraphics[scale=.5]{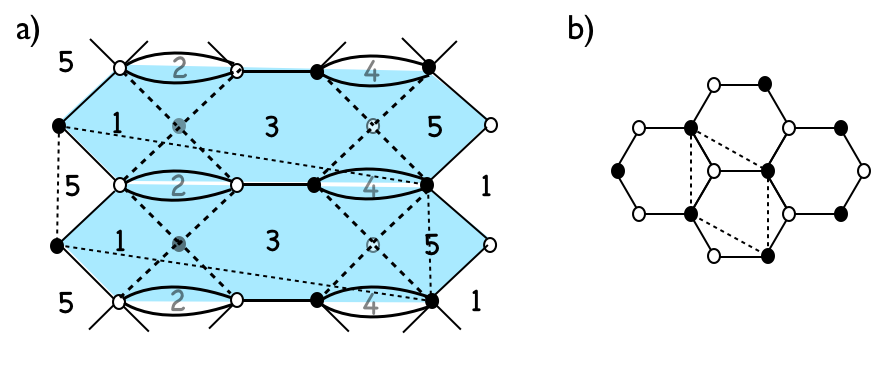} 
\caption{\small a) The confinement and chiral symmetry breaking of the gauge factors 2 and 4 can be represented in the dimer diagram. b) Upon recombining the faces 1,3 and 5, and collapsing the faces 2 and 4, we are left with the dimer of $\IC^3$. The blue shade in (a) suggest the shape of the final face in (b).}
\label{fig:deform-z2w3}
\end{center}
\end{figure}

The strong dynamics of the gauge theory due to the fractional branes can be easily studied using the dimer diagram, as follows. For our purposes, it is sufficient (and later on, necessary) to take the gauge factors on faces 2 and 4 of equal rank, so both reach the infrared of their duality cascade simultaneously. Both groups confine, and their elementary flavours must be recast in terms of mesons, which moreover get non-trivial vevs and break some of the flavour symmetries to their diagonal subgroup. Concretely, the dynamics of face 2 forces the recombination of the gauge symmetries associated to faces 1 and 3, while the dynamics of 4 recombines 3 and 5. This is shown in Figure \ref{fig:deform-z2w3}a, where the confining faces are ultimately shrunk and its endpoint nodes are collapsed. The resulting gauge theory contains only one face, and its dimer diagram is described in Figure \ref{fig:deform-z2w3}b. It corresponds to the quiver gauge theory of D3-branes in flat $\IC^3$, showing that the complex defomation has smoothed out the singular geometry completely.

\subsubsection{The orientifold throat with O3-plane}

We are now ready to consider warped throats with O3-planes. As discussed above, the general strategy is to search for complex deformed geometries which admit a $\IZ_2$ symmetry flipping the sign of the holomorphic 3-form, and having isolated fixed points. Since the equations for complex deformed geometries are not always easily obtained, it is simpler to work in the holographic dual field theory, by considering orientifolds of dimer diagrams. 

They are basically obtained by modding out the dimer diagram by some $\IZ_2$ symmetry, subject to some rules which were characterized in \cite{Franco:2007ii}, to which we refer the reader for details. For our present purposes, it is crucial that the $\IZ_2$ symmetry of the dimer is compatible with the introduction of the fractional branes which trigger the duality cascade and the infrared confinement, equivalently with the complex deformation of the dual throat geometry. Otherwise, the orientifold quotient is incompatible with the complex deformation required to support the throat.

\begin{figure}[htb]
\begin{center}
\includegraphics[scale=.5]{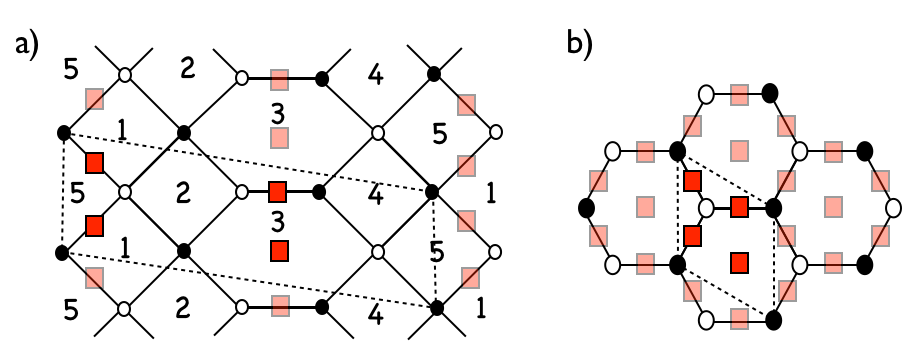} 
\caption{\small a) The orientifold of the toric geometry is represented as a symmetry of the dimer diagram. In this case we have a reflection with respect to the points signalled by a red square. For simplicity we have highlighet the squares in the unit cell, while their periodic copies are faded. The symmetry is compatible with the introduction of fractional branes, and their strong dynamics, after which we are left with an orientifold of the theory of D-branes in $\IC^3$, defined by the inherited point reflection in the dimer (b).}
\label{fig:orientifold-z2w3}
\end{center}
\end{figure}

In Figure \ref{fig:orientifold-z2w3}a we show one such $\IZ_2$ symmetry of the dimer for $xy=z^3w^2$, corresponding to a reflection with respect to the points signalled with a square. Since the $\IZ_2$ maps the face 2 to the face 4 and viceversa, it is compatible with the introduction of (an equal number of) fractional branes in these faces. Correspondingly, the strong dynamics process in Figure \ref{fig:deform-z2w3} is also $\IZ_2$ invariant, and results in an orientifold of the dimer of $\IC^3$, see Figure  \ref{fig:orientifold-z2w3}b. The geometric counterpart of these statements is that the orientifold is compatible with the complex deformation of the geometry, when the two deformation parameters are related. The result of the deformation is an orientifold of a smooth geometry, which locally is an orientifold of $\IC^3$.

We must check that the orientifold actually corresponds to an O3-plane.  As studied in \cite{Franco:2007ii},  the  action of the orientifold symmetry on the geometry (and therefore, its fixed point set) is determined by the choice of orientifold projection imposed on dimer elements fixed under the orientifold (i.e. on top of the red squares). For instance, in order to have an O3-plane (rather than an O7-plane)  on the final $\IC^3$, the orientifold must project the gauge group down to an $SO$ factor, while the fields from orientifold-fixed edges must be projected down to two-index antisymmetric representations\footnote{In the conventions of \cite{Franco:2007ii}, the fixed point in the middle of the face has sign $+$, while the other three have signs $-$. We remind the reader that these signs are mere labels and do not describe the physical charge of the resulting O3-plane (which for our choice is indeed an O3$^-$-plane).}. This same choice, inherited back in the original dimer of $xy=z^3w^2$,  guarantees that it defines an orientifold with isolated fixed points in the toric geometry.

Let us finally turn to a more explicit description of the orientifold action on the geometry, as given by the defining equation (\ref{ecuaz2w3}). The first observation is that the orientifold is a symmetry only if the deformation parameters are related, i.e. $\epsilon_1=-\epsilon_2$. Denoting any of them by $\epsilon$, the geometry reads
\beqa
xy=z (zw+\epsilon)(zw-\epsilon)
\label{orient-ecua-z2w3}
\eeqa
The holomorphic 3-form can be written as $dz\, dw\, dx/x$. The recipe in \cite{Franco:2007ii} allows to read out the orientifold action on the coordinates (by realizing them as mesons of the field theory), which agrees with the simplest guess
\beqa
x\to -x\quad,\quad y\to -y\quad ,\quad w\to -w\quad ,\quad z\to z
\eeqa
This is a symmetry of (\ref{orient-ecua-z2w3})  acting by exchange of the two pieces in parenthesis (thanks to the relation between the deformation parameters). The holomorphic 3-form is odd, as it should. Finally, the set of fixed points is given by $x=0$, $y=0$, $w=0$, and using the defining equation $z=0$. Therefore the origin is the only fixed point, and defines the location of the O3-plane.

\begin{figure}[htb]
\begin{center}
\includegraphics[scale=.5]{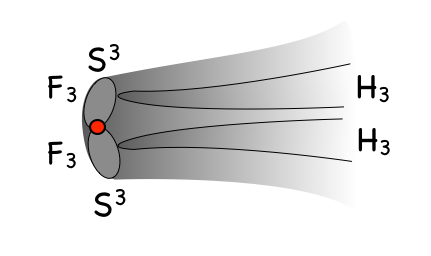} 
\caption{\small The warped throat with an O3-plane at its infrared tip, shown as a red circle.}
\label{fig:orientifold-throat}
\end{center}
\end{figure}

The field theory analysis establishes that the O3-plane sits at the bottom of the dual warped throat. Indeed, the duality cascade occurs in the gauge theory described by the orientifolded dimer, and the confinement  produces an orientifold of the smoothed out infrared geometry. In fact, this is also directly clear form the geometry, since the equality of the numbers of fractional branes leads to an equality of the number of RR flux units in the two 3-cycles, so both are located at the bottom of the throat, and so is the O3-planes, which defines the fixed point under their exchange. The structure of the warped throat is shown in Figure \ref{fig:orientifold-throat}.

The holographic dual picture also makes it manifest that these more general throats have a behaviour for the warp factor which is essentially identical to the KS throat. More concretely, if the throat is dual to a cascade of Seiberg dualities in a theory with $M$ fractional branes and $N=KM$ D3-branes (at some UV cutoff scale), the warp factor at the bottom of the throat is
\beqa
z\sim  \exp\big(-\frac{2\pi K}{Mg_s}\big)
\eeqa
Actually, different throats lead to different order 1 numerical factors in the exponent, related to the amount of D3-branes disappearing in a duality period. The important point is however that the parametric dependence in $K$ and $M$ is maintained, and therefore the throats lead to exponential suppressions with respect to the bulk or cutoff scales.

\section{Coupling the Nilpotent field to moduli and matter fields}

We have seen that the parameter M reflects the breaking of supersymmetry, and the goldstino belongs to a chiral nilpotent superfield  $X$. In this section we  provide a preliminary discussion of how $X$ might couple to the moduli and matter fields in a full string compactification, leaving a more detailed description  for the future. 

  Let us assume that the complex structure moduli and dilaton have been stabilised supersymmetrically by the fluxes, and consider as simple model of the remaining dynamics. We consider the (for simplicity, a single) K\"ahler modulus $T$, the nilpotent superfield $X$, and a chiral superfield $C$ as a representative matter field, which we assume to be stabilized at $C=0$ but we keep it in the action to study how its components split after supersymmetry breaking.

In general the K\"ahler potential can be written as 
\be
K=-3\log\left(T+T^*\right) +  c \R^n \, XX^* + Z CC^* + \cdots
\ee

where
\be
Z=\R^m +  b \R^k XX^*
\ee

The coefficients $c,b$ are arbitrary (after absorbing other coefficients as field redefinitions of $C$ ) and also the `modular weights' $n,m,k$ which   are expected to be non-positive rational numbers. Particular cases are $n,m,b=0$ corresponding to canonical kinetic terms for both $X$ and $C$. Also  the case $n=m=-1,  k=-2,  b=1/3$ corresponds to the K\"ahler potential $K=-3\log (T+T^*-  CC^*- cXX^*)$ after scaling properly the fields $C$ and $X$. The superpotential is
\be
W=W_0+ M X  + W_{\rm matter} + W_{\rm np}
\ee
where both $W_0$ and $M$ are functions of the complex structure  moduli and dilaton at their minimum, $W_{\rm matter} =C^3+\cdots$, and $W_{\rm np}=Ae^{-aT}$. We will work in the limit  $a\R\gg 1$ in order to have a proper non-perturbative expansion. 

The coupling between $T$ and $X$ modifies the appearance of $M$ in the scalar potential and gives:
\be
V_{\rm{uplift}} = \frac{|M|^2}{c \R^{n+3}}
\ee
Notice that this agrees with the KKLT expression for $n=0$ and the KKLMMT (warped) expression for $n=-1$.  In this case the warping can be absorbed in the coefficient $|M|^2/c$.

Even though without the uplifting term the field $T$ is stabilised supersymmetrically ($D_TW=\partial_TW+K_TW=0$), the presence of the uplift term induces a shift on the value of $T$ that generates a non-zero $F$ term for $T$. We find:
\be
F_T=e^{K/2} D_TW\sim \frac{3 W_0}{\R^{3/2}} \epsilon 
\ee
with 
\be
\epsilon = \frac{3+n}{c}\, \frac{1}{ a^2\R^2}
\ee
This induces, as expected, a small shift in the scalar potential:
\be
V_0= \frac{|M|^2}{c \R^{n+3}}-3m_{3/2}^2 +{\mathcal{O}}\left(\epsilon \,m_{3/2}^2\right)
\ee

A nonvanishing value of $M$ reflects the breaking of supersymmetry. However its impact on matter fields $C$ and standard model gauginos needs to be computed. We will assume here for simplicity \footnote{The validity of this assumptions will be tested when the complete supergravity models interacting with a nilpotent multiplet and other chiral multiplets will be constructed.}, that  the standard expressions for soft terms  \cite{ST} (gaugino masses $M_{1/2}$, scalar masses $m_0$ and trilinear $A$- terms) can be applied even in case that one of the superfields is nilpotent
\bea
M_{1/2}&=& \frac{1}{f+f^*}F^I\partial_I f \nonumber \\
A_{\alpha\beta\gamma} & = & F^IK_I+F^I\partial_I\log Y_{\alpha\beta\gamma}-F^I\partial_I\log\left(Z_\alpha Z_\beta Z_\gamma\right) \\
m_0^2 & = & V_0+m_{3/2}^2 - F^IF^J\partial_I\partial_J \log Z \nonumber
\eea
Here, indices $\alpha$, $\beta$, $\gamma$ label different matter fields, indices $I,J$ run over moduli fields and in our case also the $X$ field. Also, $f$ is the holomorphic gauge kinetic function of the visible sector, depending only on moduli fields and dilaton, and $Y_{\alpha\beta\gamma}$ are Yukawa couplings among matter fields. It is clear from these expressions that that the $F$ term of the nilpotent superfield $X$ only affects the scalar masses\footnote{This qualitative feature for anti-D3-brane SUSY breaking was derived in \cite{Camara:2003ku} in the probe approximation.}: first, note that $X$ is localised, so $f$ cannot depend on it, and hence the contribution of $F_X$ to gaugino masses vanishes; second, since the scalar component of $X$ vanishes in the vacuum, it gives no contribution to $A$. On the other hand the first and third terms in the expression for the scalar masses do depend on $M$. Using 
 $F^X=e^{K/2}K^{-1}_{XX^*} D_XW =M/\R^{n+3/2}$ and $F_T/\R= \mathcal{O}\left( \epsilon^{1/2} m_{3/2}\right)$ we find:
 \bea
 M_{1/2}, A &=& \mathcal{O}\left( \epsilon^{1/2} m_{3/2}\right)\nonumber\\
 m_0^2 &= & V_0+m_{3/2}^2-F^XF^{ X^*} \partial_X\partial_{ X^*}\log Z + \mathcal{O}\left( \epsilon m_{3/2}^2\right)\nonumber\\
 &=& V_0+m_{3/2}^2-\frac{b}{c^2} \R^{k-m-2n-3} |M|^2+\mathcal{O}\left( \epsilon m_{3/2}^2\right)
 \eea
  After tuning the vacuum energy to $V_0\sim 0$ we can see that the soft scalar masses are of order the maximum between the second and third term. For $k=m+n$ these are all of order $m_0^2\sim m_{3/2}^2$ as expected generically. Furthermore for $b=c/3$ these two terms combine with each other 
 to give
 \be
 m_0^2=V_0-\frac{1}{3}\left( \frac{|M|^2}{c\R^{n+3}}- 3 m_{3/2}^2 \right)+ \mathcal{O}\left( \epsilon m_{3/2}^2\right) = \frac{2}{3} V_0 + \mathcal{O}\left( \epsilon m_{3/2}^2\right)
 \ee
 In this case the leading order contribution to all soft terms comes from $F_T$. Since at the minimum 
 $a\R\sim \log W_0$ and all scales are measured in units of $M_{\rm planck}$, this implies the soft terms are of order $m_{soft}\sim m_{3/2}/ \log\left(\frac{M_{planck}}{m_{3/2}}\right)$.  Notice that a K\"ahler potential of the form $K=-3\log\left( T+T^*-CC^*-XX^*\right)$ satisfies all these conditions and therefore in this case all soft terms will be subdominant with respect to the gravitino mass and other effects such as anomaly mediation should also be considered. This result agrees with the proposal of \cite{choi} for soft terms in the KKLT scenario. A similar cancelation also occurs in the sequestered `ultra-local' case in the LARGE volume scenario \cite{sequestered}.
 
 In summary we can see that both gaugino masses and trilinear $A$ terms are  suppressed with respect to the gravitino mass as $m_{3/2} /\log\left(\frac{M_{planck}}{m_{3/2}}\right)$. On the other hand,  the scalar masses depend on the precise form of the matter K\"ahler potential. In some models, they  are of order $m_{3/2}$ (or larger), in which case we will have a realization of split supesymmetry. If instead  the cancelation mentioned above occurs, then they are also of order  $m_{3/2} /\log\left(\frac{M_{planck}}{m_{3/2}}\right)$.  Further studies of these issues are left for future work.

\section{Discussion}

It is rather unusual to encounter a supergravity model with a nilpotent multiplet. In fact, such a complete supergravity model action with explicit spontaneously broken local supersymmetry was not even presented in the literature.  Some partial  work in this direction includes  \cite{Deser:1977uq} where a proposal was made how to generalize the global Volkov-Akulov model to a locally supersymmetric one and \cite{Volkov:1994vg}, where the curved superspace formulation of the VA goldstino theory was studied. 
However, in both cases the action as well as the local supersymmetry rules were not presented in a complete form. A related  work was performed in \cite{Dudas:2000nv} where  the leading order d=10  Lagrangian with the coupling of gravitino to  Volkov-Akulov model was studied  (see also \cite{Pradisi:2001yv} for follow-up work.). A d=4
supergravity with nilpotent multiplet is not available in the textbooks and the complete fermion part of it is not known.

The fact that the globally supersymmetric VA action in the form \rf{VA} as well as in the form \rf{VA1} has a negative constant $-M^2$ in the action is well known. However, only when VA goldstino has a consistent coupling to gravity, this term in the action becomes $-\sqrt {-g} \, M^2$ and indicates a contribution to a positive cosmological constant. It was shown in \cite{Kachru:2003sx} that when anti-D3-brane is coupled to d=10 supergravity compactified to d=4, indeed, such a term is present and leads to a KKLT uplift. A generalization of this argument  in the setting where also fermions are present on the world-volume of the anti-brane was given in 
\cite{Bergshoeff:2015jxa}. However, the supersymmetry on the 
anti-D3-brane on top of the O3-plane still has a global nature, see eq. \rf{eq:lambda} where $\zeta$ is the space-time independent fermionic parameter.

To discuss the issue of a positive cosmological constant, one would like to have also an action with  local supersymmetry where the positive contribution to the energy comes from the potential. The corresponding action together with the local spontaneously broken supersymmetry  is now under construction \cite{WP}.

It is therefore very interesting that the string theory with open strings at the 1-loop level and closed string at the tree level provides a configuration which is precisely the one with fermions on a single  anti-D3-brane on top of the  O3-plane. In this paper we have given a detailed computation which explains the origin of the nilpotent chiral multiplet in string theory. The computation of the open string spectrum supports the use of the spontaneously broken supersymmetry with a fermionic goldstino multiplet.  

We have also studied possibilites to place an orientifold O3-plane at the tip of the throats. It turned out that  the familiar KS model \cite{Klebanov:2000hb} does not admit orientifold actions with isolated fixed points, and therefore it cannot host an O3-plane at its bottom. However, it is easy to find general constructions of other throats, based on deformations of slightly more general toric geometries, which have the same features concerning the generation of warp factors by fluxes supported on 3-cycles, yet admit orientifold  actions with fixed points at the bottom of the throat. We have provided  one such example, based on an orientifold of a generalized conifold, and phrased its detailed description in the terms of the holographic dual quiver gauge field theory. The use of dimer diagrams allows a exquisite tracking of both the field theory RG flow, which involves a Seiberg duality cascade triggered by a precise set of fractional branes, and the infrared confinement phenomena dual to the IR capping of the throat by the 3-cycles.

We have also shown that the spectrum of the nilpotent goldstino can be realized, below the flux scale, in one $\aD3$-brane on top of an O7-plane, and that this construction can be easily implemented in the familiar KS throat (which does admit O7-planes stretching along the radial direction down to the bottom of the throat).

More studies of the relation to string theory would be desirable with regard to phenomenological cosmological models, involving the nilpotent multiplet and  describing inflation with supersymmetry breaking and cosmological constant in the vacuum.

An important aspect of this story is the following. The mysterious origin of dark energy, which is well presented by the positive cosmological constant at present, might be due to spontaneously broken local supersymmetry, like the Higgs effect is due to spontaneously broken gauge symmetry. A tiny cosmological constant results from an incomplete cancellation of the positive goldstino and negative gravitino 
contribution to the supergravity energy. Still the argument for the smallest of $\Lambda$ has to rely on the existence of the landscape. In our paper we studied possibilities to get many different values of a positive contributions to cosmological constant due to many choices of fluxes integers $M$ and $K$ such that   the SUSY breaking parameter is $\sim  \exp\big(-\frac{2\pi K}{Mg_s}\big)$.

There are several open questions that are left for future work. On the formal side, it would be interesting to device a string description from which to derive the nilpotency of the goldstino multiplet more directly. In the model building direction, it would be desirable to realize the embedding of our local constructions in concrete compact Calabi-Yau orientifolds, even though no fundamental obstacles are foreseen,  in order to have a fully global realization of this de Sitter scenario and address global issues such as explicit implementation of inflation in concrete models.

 Finally, the concrete realization of supersymmetry breaking in terms of nilpotent superfields opens up the possibility to analyze the structure of soft supersymmetry breaking terms for standard model fields in a very explicit way. We started this discussion in section 4 but there are several open questions.
 Determining the precise form of the K\"ahler potential for the nilpotent and matter fields is needed in order to extract more concrete phenomenology. Studying the generic case with more K\"ahler moduli would be interesting to consider. Also sequestered scenarios of supersymmetry breaking as those described in \cite{sequestered} are sensitive to the concrete uplifting mechanism and could be approached using the nilpotent goldstino superfield discussed here. Our results give further motivation for addressing these questions.

\section*{Acknowledgments}

We are grateful  to  L. Aparicio, E. Bergshoeff,  J.~J.~M.~Carrasco, M. Cicoli, K. Dasgupta, S. de Alwis, S. Ferrara, S. Franco, D. Freedman, E. Dudas,  M. Goodsell, M. Gra\~na, L. Ib\'a\~nez, S. Kachru, A. Linde, L. McAllister, J. Polchinski, A. Puhm, A. Retolaza, D. Roest,  F. Zwirner,  A. Van Proeyen and T. Wrase  for stimulating  discussions.  The work of RK
was supported by the SITP,   by the NSF Grant PHY-1316699 and  by the Templeton foundation grant `Quantum Gravity Frontiers'. AU is partially supported by the grants  FPA2012-32828 from the MINECO, the ERC Advanced Grant SPLE under contract ERC-2012-ADG-20120216-320421 and the grant SEV-2012-0249 of the ``Centro de Excelencia Severo Ochoa" Programme.
We are grateful to the organizers and participants of the String Phenomenology 2015 workshop where this work was initiated and those of the Strings 2015 and PASCOS 2015 that facilitated the completion of this project.

\end{document}